\documentclass{aa}  
\usepackage{graphicx}
\usepackage[varg]{txfonts}

\usepackage{txfonts}
\usepackage{subcaption}         
\usepackage{lscape}             
\usepackage{placeins}           
                                
\usepackage{natbib}

\usepackage{blindtext}
\usepackage{bm}
\usepackage{amsmath}
\usepackage{xcolor}

\bibpunct{(}{)}{;}{a}{}{,}

\newcommand{\correct}[1]{{\color{black}#1}}

\def\deg{\hbox{$^\circ$}}

\def\utw{\ensuremath{\smash{\rlap{\lower5pt\hbox{$\sim$}}}}}
\def\udtw{\ensuremath{\smash{\rlap{\lower6pt\hbox{$\approx$}}}}}

\begin{document}

\nolinenumbers

\title{\correct{1.5D investigation of the Hanle effect in the Ca {\sc i} 4227 line using partial frequency redistribution}}

   \subtitle{\correct{Comparison of synthetic Stokes profiles in Bifrost and MURaM-ChE rMHD models}}
  \author{Devang Agnihotri\inst{1,2}        
        \and L. S. Anusha\inst{1,2}
        \and D. Przybylski\inst{3}
        \and R. H. Cameron\inst{3}
        \and S. K. Solanki\inst{3,4}
        }
   \institute{Indian Institute of Astrophysics, II Block, Koramangala, Bengaluru 560 034, India
   \and Pondicherry University, R.V. Nagar, Kalapet, Pondicherry 605014, UT of Puducherry, India
   \and Max Planck Institute for Solar System Research, Justus-von-Liebig Weg - 3, Goettingen 37077, Germany 
   \and School of Space Research, Kyung Hee University, Yongin, Gyeonggi 17104, Republic of Korea}

\abstract
{
Modeling scattering polarization in the Ca\,{\sc i} 4227 \AA\, line is important for diagnosing chromospheric magnetic fields. This requires solving the polarized radiative transfer equation including the Hanle effect and partial frequency redistribution (PFR).
Plane-parallel, semi-empirical models are known to be inadequate for accurately reproducing the observed scattering polarization profiles of this line.
}
{
The primary aim of this work is to investigate the relative influence of heights of formation and magnetic fields (via Hanle effect) in realistic solar model atmospheres on the emergent spatially averaged synthetic Stokes profiles of the Ca {\sc i} 4227 \AA\, line.
}
{
For this purpose, we use one-dimensional vertical columns derived from snapshots of recent three-dimensional radiative magnetohydrodynamic (rMHD) simulations of the solar chromosphere performed with the Bifrost and MURaM-ChE codes as model atmospheres, and carry out 1.5D non-local thermodynamic equilibrium polarized radiative transfer calculations including PFR and the Hanle effect.
}
{
The intensity profiles from both simulations coincide at the line core whereas the line wings in MURaM-ChE show higher intensity as these signals originate from deeper, hotter regions. 
The polarization signals are formed at greater heights in MURaM-ChE than in Bifrost.
In the nonmagnetic case, the emergent, spatially averaged profiles from Bifrost exhibit a larger polarization amplitude in the line core, whereas MURaM-ChE ones shows enhanced polarization in the wings.
When the impact of magnetic fields from the 3D rMHD simulations is included, the spatially averaged profiles indicate stronger Hanle depolarization of $\langle Q \rangle / \langle I \rangle$ in the Bifrost snapshot.
}
{
Because scattering polarization is highly sensitive to the heights at which the polarization signals are formed, the differing thermal and magnetic structures at the corresponding line formation heights in the two simulations lead to distinct Stokes profiles.
In particular, the Hanle depolarization signatures indicate that the line-core formation region in the MURaM-ChE snapshot is less strongly  magnetized than in the Bifrost snapshot. 
%
}
\maketitle
\nolinenumbers
\section{Introduction}
\label{sect:intro}

The Ca {\sc i} 4227 \AA\, line  exhibits the largest scattering polarization in the visible spectrum when observed near the solar limb. 
%
These polarization signals originate from resonance scattering \citep[][]{chandrasekhar1950} of anisotropic radiation by Ca {\sc i} atoms in the solar atmosphere. 
The presence of weak magnetic fields modifies the line-core of these scattering polarization signals due to the Hanle effect \citep[see, e.g.,][]{stenflo82}. This implies that the Ca {\sc i} 4227 \AA\,line can be used to deduce weak magnetic fields in the lower to middle solar chromosphere \citep[e.g.,][]{1998_Bianda}. 
%

%
Observations of the linear polarization of the Ca {\sc i} 4227 \AA\, line show a typical triple peak structure \citep[][]{mf92, 2005_Holzreuter} caused by scattering under partial frequency redistribution \citep[PFR, see e.g.,][]{mihalas_book_2015}.
Magnetic field diagnostics based on the Hanle effect in solar spectral line polarization require the computation of Stokes profiles by solving the non-local thermodynamic equilibrium (non-LTE) polarized radiative transfer equation (PRTE) in model solar atmospheres in the presence of weak magnetic fields.

Various studies focused on modeling the Ca {\sc i} 4227 \,\AA\, line using one-dimensional (1D) semi-empirical  model solar atmospheres.
E.g., in  \citet{Anusha_2011} forward scattering polarization observations taken near disk center were modeled, while center to limb variations were studied in \citet{2014_supriya} and \citet{2015_supriya} using FAL model atmospheres \citep[see][]{Fontenla_1993, avrett_1995}. 
These studies used angle-averaged PFR in their modeling.
\citet{2021_janett} used the 1D FAL-C model atmosphere to compare the angle-dependent and the angle-averaged PFR  effects in the line polarization. 
Magneto-optical effects on Stokes profiles were studied by \citet{alsina_2018}. 
These effects were incorporated by \citet{emilia_2022} 
who attempted to calculate the line polarization angle and the volume filling factor for magnetic fields using the FAL-C model atmosphere.
%
%
Recently \citet{2024_beluzzi} studied the effect of complete frequency redistribution approximation \citep[CFR, see][]{mihalas_book_2015} with respect to angle-averaged and angle-dependent PFR on the center to limb variation of scattering polarization in this line using arbitrary magnetic fields and the 1D FAL-C model atmosphere.

While 1D models are helpful for a qualitative comparison with observed Stokes profiles, they  
produce profiles discrepant from the observed ones \citep[see, e.g.,][]{mf92,LSA_2010}. 
A slight modification of the temperature profiles in the FAL atmospheres helped reducing the discrepancies as found in \citet{2014_supriya} and \citet{2015_supriya}. 
Further, the semi-empirical model atmospheres do not contain information about the magnetic fields in the solar atmosphere.
Thus, their utility is limited  to the use of arbitrary magnetic fields for modeling the Hanle effect in spectral line polarization.

Three-dimensional (3D) radiative magnetohydrodynamic (rMHD) simulations enable a more realistic representation of the solar atmosphere by accurately modeling its complex structures and dynamic behavior. 
While simulations of the solar photosphere are well established as generated, e.g., by \citet{1998_sn,2005_voegler}, modeling observed polarization in resonance lines such as the Ca {\sc i} 4227 \AA\, demands realistic representations of the solar chromosphere. 
The 3D rMHD codes Bifrost \citep[][]{2011_gud} and MURaM \citep[][]{2005_voegler} have been extended to treat the chromosphere, as detailed in \citet{2016_bifrost} and \citet{2022_muram}, respectively. 

With these advances in development of the 3D rMHD models of the solar chromosphere, several efforts have been devoted to studying the Ca\,{\sc i} 4227 \AA\ scattering polarization profiles.
E.g., \citet[][]{Mathur_2025} studied the scattering polarization in the Ca {\sc i} 4227 \AA\ line using the PRTE with angle-averaged PFR in 1D vertical columns extracted from a snapshot of Bifrost 3D rMHD simulations (the so-called 1.5D approximation), whereas \citet{2021_bestard} examined the effect of horizontal plasma bulk velocity gradients in these simulations on the polarization of this line using 3D PRTE calculations with CFR.
\citet[][]{riva_2023} and \citet[][]{guerreiro_2023} explored the effect of angle-dependent PFR and plasma bulk velocities on scattering polarization of this line using a few sample 1D vertical columns from the Bifrost simulations. The spatiotemporal evolution of Hanle and Zeeman synthetic polarization was studied in \citet {carlin_2017} also using Bifrost simulations.

In this paper our aim is to study and compare the formation of the synthetic 
Stokes profiles of the Ca {\sc i} 4227 \AA\, line in snapshots of 3D rMHD solar chromospheric simulations.
To this end, we extract 1D vertical columns from snapshots of 3D rMHD simulations by Bifrost \citep[][]{2016_bifrost} and MURaM-ChE \citep[][]{2022_muram} and solve the PRTE in 1.5D  on these columns including PFR.
%
Additionally, we investigate the impact of magnetic field in the two simulations on the line core polarization via the Hanle effect.

In Section~\ref{sec:theory}, we summarize the polarized line formation theory for a two-level atom with unpolarized ground state, which we used in this paper. 
In Section~\ref{sec:methods}, the numerical methods employed in this work and the relevant computational details are given. 
The results are presented in Section~\ref{sec:results}.
Finally,  a summary of our findings and corresponding discussions are presented in Section~\ref{sec:conclusions}.

\section{Theory and radiative transfer formulation}
\label{sec:theory}
The Stokes profiles are computed by solving the non-LTE PRTE in irreducible spherical tensor formalism \citep[][]{2007_frisch}. The relevant theory is already presented in \citet{Anusha_2011}, here we present a summary for the sake of completeness. 

 \subsection{Stokes parameter formulation}
 \label{stokes_formulation}
 The Stokes vector PRTE for a two-level atom with unpolarized ground level, in a 1D planar medium with weak magnetic fields and without plasma bulk velocities can be written as

\begin{equation}
\begin{split}
\mu\frac{\partial{\textrm{\bf{I}}}(\lambda,\boldsymbol{\Omega},z)}{\partial z} = -[\kappa_{l}(z)\phi(\lambda,z)+\kappa_{c}(\lambda,z)+\sigma_{c}(\lambda,z)]\\
\times [\textrm{\bf{I}}(\lambda,\boldsymbol{\Omega},z)-\textrm{\bf{S}}(\lambda,\boldsymbol{\Omega},z)],
\end{split}
\end{equation}

\noindent where $\textrm{\bf{I}} = (I,Q,U)^{T}$ is the Stokes vector and $\textrm{\bf{S}}= (S_{I},S_{Q},S_{U})^{T}$ the source vector, $\kappa_l$ is the wavelength-integrated line
absorption coefficient, while $\sigma_c$ and $\kappa_c$ are the continuum scattering and continuum absorption coefficients, respectively.
The total opacity coefficient for the Ca {\sc i} 4227 \AA\, line and the continuum is $\kappa_{\textrm{total}}(\lambda,z) = \kappa_l(z)\phi(\lambda,z) + \sigma_c(\lambda,z)+\kappa_c(\lambda,z)$.
Here, $\lambda$ is the wavelength, $z$ is the vertical coordinate, and $\Omega=(\theta,\varphi)$ defines the direction of the ray where $\theta$ is the polar angle (with {$\mu=\cos\theta$}) while $\varphi$ is the azimuth angle.
$\phi(\lambda,z)$ is the Voigt profile with damping parameter $a = {\Gamma_{\textrm{total}}}/{(4\pi\Delta\nu_{D}})$.
Here, $\Gamma_{\textrm{total}}= \Gamma_R+\Gamma_E$ with $\Gamma_R$ and $\Gamma_E$ being the radiative de-excitation rate and elasic collision rate respectively.
%
%
%
The Doppler width is given by $\Delta\nu_D = \lambda_0^{-1}\sqrt{2k_BT/M_a + v_{\mathrm{turb}}^2}$, where $M_a$ is the mass of the atom, $T$ is temperature, $k_B$ is the Boltzmann constant, $v_{\textrm{turb}}$ is the micro-turbulent velocity, and $\lambda_0$ is the line center wavelength. 
The  
Stokes $V$ component is neglected here because it is not generated by the weak-field Hanle effect considered in this paper. 
This assumption also implies the neglect of the magneto-optical effects that can impact the wing polarization \citep[see][]{alsina_2018}.
This is consistent with the objective of the present work, which is not to model observed Stokes profiles, but rather to investigate the formation of scattering polarization profiles in rMHD model atmospheres.
Further, in this work we do not include plasma bulk velocities, as our primary objective is to investigate the relative influence of model atmospheres derived from two different 3D rMHD simulation codes on the formation of scattering polarization profiles in the presence of magnetic fields.\\
The motivation for these approximations stems from the relative lack of studies that address this problem using rMHD model atmospheres under such assumptions. 
We emphasize, however, that the inclusion of plasma bulk velocities is expected to significantly impact  the polarization profiles presented here \citep[see e.g.,][]{guerreiro_2023}. 

Inclusion of plasma bulk velocity fields requires the scattering physics to take angle-dependent PFR into account for accurate calculations. %
However, the locally co-moving approach of \citep[][]{mvn2002, Leenaarts2012} allows the use of angle-averaged PFR as an approximation, which is a less restrictive assumption than completely neglecting plasma bulk velocities. %
In our case, the current implementation of the POLY code is restricted to static atmospheres and the use of angle-averaged PFR. %
The incorporation of plasma bulk velocities together with angle-dependent PFR is currently in progress and will be addressed in a forthcoming paper.

In a two-level atom model with an unpolarized ground level, the source vector is defined as

\begin{eqnarray}
\label{stot}
&&\textrm{\bf{S}}(\lambda,\boldsymbol{\Omega},z) = \frac{\kappa_l(z) \phi(\lambda,z)\textrm{\bf{S}}_{\ell}(\lambda,\boldsymbol{\Omega},z)}{\kappa_{\textrm{total}}(\lambda,z)} \\ \nonumber
&& +\frac{\sigma_c(\lambda,z)\textrm{\bf{S}}_c(\lambda,\boldsymbol{\Omega},z)+\kappa_c(\lambda,z)B_{\lambda}(z)\textrm{\bf{U}}}{\kappa_{\textrm{total}}(\lambda,z)}.
\end{eqnarray}
Here, $\textrm{\bf{U}}$ = $(1,0,0)^T$ and $B_{\lambda}$ is the Planck function.
The line source vector can be written as 

\begin{equation}
\begin{split}
\textrm{\bf{S}}_{\ell}(\lambda,\boldsymbol{\Omega},z)=\epsilon B_{\lambda}(z)\textrm{\bf{U}}+\int_{-\infty}^{+\infty}\oint\frac{\hat{R}(\lambda,\lambda',\boldsymbol{\Omega},\boldsymbol{\Omega'},z,\textbf{B})}{\phi(\lambda,z)}\\
\times \textrm{\bf{I}}(\lambda',\boldsymbol{\Omega'},z)\frac{d\boldsymbol{\Omega'}d\lambda'}{4\pi}.
\end{split}
\end{equation}
\noindent
Here, $\hat{R}$ is the Hanle redistribution matrix 
 \citep[approximation III of][]{1997aBommier,1997bBommier} which uses angle-averaged PFR functions. 
 \textrm{The} magnetic field \textbf{B} is extracted from the 3D rMHD simulation snapshots. %
 The thermalization parameter is defined as $\epsilon$ = ${\Gamma_I}/(\Gamma_R+\Gamma_I)$, where $\Gamma_I$ is the inelastic collision rate.
The continuum source vector is

\begin{equation}
\textrm{\bf{S}}_c(\lambda,\boldsymbol{\Omega},z)=\oint \hat{P}(\boldsymbol{\Omega},\boldsymbol{\Omega'})\textrm{\bf{I}}(\lambda,\boldsymbol{\Omega'},z)\frac{d\boldsymbol{\Omega'}}{4\pi}{,}
\end{equation}
with $\hat{P}$ being the Rayleigh scattering phase matrix, as frequency coherence is assumed for the continuum.
Also, primed quantities denote incoming photons, while unprimed ones are for outgoing photons (after scattering).

\subsection{Spherical irreducible tensor decomposition}
\label{spherical tensors}
The Stokes source vector \textbf{S} and the Stokes vector \textbf{I} can be represented in the form of spherical tensor components as ${\bm{\mathcal{I}}}$ and ${\bm{\mathcal{S}}}$ whose elements are $I^{K}_{Q}$ and $S^{K}_{Q}$, introduced by \citet[][]{2007_frisch}, with $K=0,2$ and $Q$ $\in$ $[-K, +K]$.
In this representation, ${\bm{\mathcal{S}}}$ becomes independent of the beam direction and ${\bm{\mathcal{I}}}$ is independent of the azimuthal angle. The PRTE can be rewritten as
\begin{equation}
\mu\frac{\partial{\bm{\mathcal{{I}}}}(\lambda,\mu,z)}{\partial z}  = -\kappa_{\textrm{total}}(\lambda,z)[{\bm{\mathcal{I}}}(\lambda,\mu,z)-{\bm{\mathcal{S}}}
(\lambda,z)], 
\label{rte-ikq}
\end{equation}
where the irreducible source vector is
\begin{equation}
\begin{split}
\bm{\mathcal{S}}(\lambda,z) = \frac{\kappa_l(z)\phi(\lambda,z)\bm{\mathcal{S}}_{\ell}(\lambda,z)}{\kappa_{\textrm{total}}(\lambda,z)}\\
+\frac{\sigma_c(\lambda,z){\bm{\mathcal{S}}}_c(\lambda,z)+\kappa_c(\lambda,z)B_{\lambda}(z){\bm{\mathcal{U}}}}{\kappa_{\textrm{total}}(\lambda,z)}.
\end{split}
\end{equation}
The line source vector $\bm{\mathcal{S}}_{\ell}$ becomes
\begin{eqnarray}
\label{stotirr}
&& {\bm{\mathcal{S}}}_{\ell}(\lambda,z)=\epsilon B_{\lambda}(z){\bm{\mathcal{U}}} \nonumber \\
&&+\frac{1}{2}\int_{-\infty}^{+\infty}\int_{-1}^{+1}\frac{\hat{\mathcal{R}}(\lambda,\lambda',z,\textbf{B})}{\phi(\lambda,z)}
\hat{\bm{\Psi}}(\mu')
\times {\bm{\mathcal{I}}}(\lambda',\mu',z)d\lambda'd\mu'.\nonumber \\
\label{line-skq}
\end{eqnarray}
$\bm{\mathcal{U}} = (1,0,0,0,0,0)^T$, $\hat{\mathcal{R}}$ is the angle-averaged PFR matrix for resonance scattering in the irreducible basis \citep[see e.g.,][]{LSA3D2011}, while ${\hat{\bm{{\Psi}}}}$ is the Rayleigh scattering phase matrix in the irreducible basis \citep[see e.g.,][]{2007_frisch}.
We define the irreducible source vector ${\bm{\mathcal{\tilde{S}}}}$ as

\begin{equation}
\label{stilde}
{\bm{\mathcal{\tilde{S}}}}(\lambda,z)=\frac{1}{2}\int_{-\infty}^{+\infty}\int_{-1}^{+1}\frac{\hat{\mathcal{R}}(\lambda,\lambda',z,\textbf{B})}{\phi(\lambda,z)}
\hat{\bm{\Psi}}(\mu')
\times {\bm{\mathcal{I}}}(\lambda',\mu',z)d\lambda'd\mu'. 
\end{equation}
The elements of $\bm{\mathcal{\tilde{S}}}$  are denoted by $\tilde{S}^{K}_{Q}$. The first element $\tilde{S}^{0}_{0}$ is the
unpolarized line source component without the thermal source contribution and the second element $\tilde{S}^{2}_{0}$ represents the linearly
polarized component of the line source vector.

The continuum scattering source vector in the irreducible basis is

\begin{equation}
{\bm{\mathcal{S}}}_c(\lambda,z) = \frac{1}{2}\int_{-1}^{+1}{\hat{{\bm{{\Psi}}}}(\mu'){\bm{\mathcal{I}}}}(\lambda,\mu',z)d\mu'.
\end{equation}
Denoting the total optical depth scale as $d\tau_{\lambda}$ = $- \kappa_{\textrm{total}}(\lambda,z)dz$ and total optical depth at the bottom of the atmosphere at a given wavelength $\lambda$ as $T_{\lambda}$, 
the formal solution of Eq.~(\ref{rte-ikq}) for $\mu>0$ can be written as

\begin{equation}
\begin{split}
{\bm{\mathcal{I}}}(\lambda,\mu,\tau_{\lambda}) = {\bm{\mathcal{I}}}_0(\lambda,\mu,T_\lambda)\exp\left(-\frac{T_\lambda-\tau_\lambda}{\mu}\right)\\
+ \int_{\tau_\lambda}^{T_\lambda}\exp\left(-\frac{\tau'_\lambda-\tau_\lambda}{\mu}\right){\bm{\mathcal{S}}}(\lambda,\tau'_\lambda)\frac{d\tau'_\lambda}{\mu},
\end{split}
\label{ikq-mup}
\end{equation}
and for $\mu<0$ the solution is

\begin{equation}
\begin{split}
{\bm{\mathcal{I}}}(\lambda,\mu,\tau_{\lambda}) = {\bm{\mathcal{I}}}_0(\lambda,\mu,0)\exp\left(-\frac{\tau_\lambda}{\mu}\right)\\
-\int_{0}^{\tau_\lambda}\exp\left(-\frac{\tau'_\lambda-\tau_\lambda}{\mu}\right){\bm{\mathcal{S}}}(\lambda,\tau'_\lambda)\frac{d\tau'_\lambda}{\mu}.
\label{ikq-mun}
\end{split}
\end{equation}
\noindent
It is assumed here that no radiation is coming into the medium from the upper boundary while at the bottom of the medium, LTE prevails; thus, we have ${\bm{\mathcal{I}}}_0=(0,0,0,0,0,0)^T$ for $\tau_\lambda$ = 0 and $\mu<0$.
While at $\tau_\lambda$ = $T_\lambda$ and $\mu>0$,  ${\bm{\mathcal{I}}}_0(\lambda,\mu,T_\lambda)$ = $(B_\lambda(T_\lambda),0,0,0,0,0)^T$.

We recall that in literature the approximation of the angle-averaged PFR refers to the combination of angle-averaged $R_{II}$ and $R_{III}$ redistribution functions according to approximation III of \citet[][]{1997aBommier,1997bBommier}. 
However, the evaluation of the $R_{III}$ redistribution function is computationally expensive, particularly in modeling efforts where the redistribution functions must be computed at every depth point in the atmosphere. 
We have verified the well known result that Stokes profiles computed using the combination of $R_{II}$ and $R_{III}$ as well as those computed using the combination of $R_{II}$ and CFR for $R_{III}$ produce visually identical Ca\,{\sc i} 4227 \AA\ Stokes profiles \citep[see][]{mihalas_1978,riva_2023}.
Therefore, owing to computational constraints, we adopt $R_{II}$ together with CFR for $R_{III}$ throughout the remainder of this paper, and refer to this configuration simply as the PFR case.
\subsection{Contribution functions}

To interpret the emergent spectra it is important to understand the height where most photons escape at specific wavelengths, which is simply referred to as the height of formation. 
This height is the spatial location $z$ in the atmosphere where the contribution function $C_I$ (defined below) is maximum. 
$C_I$ is defined \citep[see][]{magain1986} by
\begin{eqnarray}
\label{ci}
&& C_I(\lambda,z,\mu)=\mu^{-1}\ln(10)\tau(\lambda_\textrm{ref},z)\frac{\kappa_{\textrm{total}}(\lambda,z)}{\kappa_{\textrm{total}}(\lambda_\textrm{ref},z)} \nonumber \\
&& \times S^{0}_{0}(\lambda,z)\exp(-\tau(\lambda,z)/\mu).
\end{eqnarray}
Here, $\lambda_\textrm{ref}$ is the reference wavelength. 
The equation, for a given $\mu$, describes contribution of source function  from optical depth $\tau(\lambda_\textrm{ref},z)$ to emergent intensity at a wavelength $\lambda$.
Similarly, we can define the contribution function for Stokes $Q$ as $C_Q$ \citep[][]{2005_Holzreuter} by replacing $S^{0}_{0}$ in Eq.~(\ref{ci}) with $S^{2}_{0}$ that contributes dominantly to Stokes $Q$ namely,
\begin{eqnarray}
\label{cq}
&& C_Q(\lambda,z,\mu)=\mu^{-1}\ln(10)\tau(\lambda_\textrm{ref},z)\frac{\kappa_{\textrm{total}}(\lambda,z)}{\kappa_{\textrm{total}}(\lambda_\textrm{ref},z)} \nonumber \\
&& \times S^{2}_{0}(\lambda,z)\exp(-\tau(\lambda,z)/\mu).
\end{eqnarray}
It is worth noting that, in 3D rMHD atmospheres, the temperature can exhibit localized spikes. 
In a few atmospheric models from our sample, this results in a dominant contribution from thermal source function at the heights where line center photons are formed. 
Since the primary objective of this paper is to investigate the scattering polarization in the line, we ignore the continuum source function and consider only the line source function for the computation of contribution function. The thermal contribution to the line source function is, however, fully retained.

\section{Computation of Stokes profiles}
 \label{sec:methods}
  \subsection{3D rMHD simulations}
 \label{subsec:1.5D}
In this paper, we use snapshots from two solar chromospheric 3D rMHD simulations, namely Bifrost and MURaM-ChE.
We use the publicly available Bifrost snapshot \citep[][]{2016_bifrost} and a snapshot from MURaM-ChE \citep[][]{2022_muram} that has an enhanced network environment similar to the Bifrost snapshot that we used.
We note here that the MURaM-ChE simulation snapshot used here is the same as the one used in \citet[][]{Ondratschek}. 
In both Bifrost and MURaM-ChE simulations non-equilibrium hydrogen ionization effects are taken into account, while radiative rates are approximated using pre-tabulated data.
Both the MURaM-ChE and Bifrost models contain a similar large scale bipolar magnetic field, representing a strong network region. 
In the Bifrost model, the bipolar field was added to a hydrodynamic simulation. 
In the MURaM-ChE model, zero net-flux magnetic fields generated by the small-scale dynamo were present before the bipole was added. 
The simulation boxes from Bifrost and MURaM-ChE snapshots span $24 \times 24$ $\textrm{Mm}^2$ horizontally and the vertical grid extends up to 14.4 Mm above photosphere.
The two codes vary in grid spacing, the vertical resolution in MURaM-ChE is uniform at $20~\mathrm{km}$, while the Bifrost model has a stretched grid.
%
%
The vertical grid spacing between the two models is similar in the photosphere and chromosphere, while the MURaM-ChE model has a higher resolution in the convection zone and corona. 
%
%
The Bifrost 3D simulation snapshot has $504\times504\times496$ grid points with uniform horizontal spacing of 48 km while MURaM-ChE has $1024\times1024\times800$ grid points with horizontal spacing of 23.4 km (twice as high).  
The combination of higher resolution and reduced numerical diffusivity in the MURaM-ChE code leads to a significantly more turbulent convection zone. This leads to a more efficient enhance the magnetic field by the small scale dynamo giving a magnetic field strength at the photosphere roughly twice that of the Bifrost model.
Further simulation details can be found in the respective papers cited above.

\subsection{Extraction of 1D vertical columns from 3D rMHD simulation snapshots}
\label{chopping}
From the 3D rMHD simulation snapshots, we extract 1D vertical columns at each grid point along the horizontal $x$ and $y$ directions. 
These vertical columns are treated as independent 1D model atmospheres with physical parameters that vary with height ($z$).
For our purpose we extract temperature, mass density, free electron density, 6 level hydrogen population densities and magnetic field parameters from these simulation snapshots.
Using the {\sc {RH}} \citep{Uitenbroek_2001} code, we compute the optical depth at 5000~\AA, denoted by $\tau_{5000}$. 
With Ca {\sc i} as the active atom, we also compute the line center optical depth at 4227~\AA, denoted by $\tau_{4227}$.

Based on $\tau_{5000}$ and $\tau_{4227}$, we construct the 1D atmospheres as follows. 
The bottom of the atmosphere is set at 500~km below the height where $\tau_{5000}=1$ (which defines $z=0$), while the top of the atmosphere is chosen at the height where $\tau_{4227} \approx 5\times10^{-4}$. 
All grid points lying outside this range are discarded for computational efficiency.
As a result, the total number of vertical grid points differs among the constructed 1D atmospheres, ranging between 163 to 861 depth points.
Further, to reduce the cost of computing Stokes profiles we select every 20th and every 40th column for Bifrost and MURaM-ChE respectively, along  both $x$ and $y$ directions.
This selection preserves the total horizontal plane spanned by the atmospheres. 
 %
To summarize, we now have a total of $26 \times 26$ 1D vertical columns for each of Bifrost and MURaM-ChE, covering a horizontal area of $24 \times 24\ \mathrm{Mm}^2$ with nearly the same horizontal grid resolution.
\subsection{Method of Solution of the PRTE}
\label{subsec:nsolution_method}
The PRTE discussed in Section~\ref{stokes_formulation} is solved in two steps. 
In the first step, we solve the multi-level unpolarized radiative transfer equation using the RH code. 
%
%
The atmospheres selected as described in Section~\ref{chopping} are used as the model atmospheres in RH code, with 1110 total wavelength points, 10 $\mu$ points in the interval $\mu \in [-1,1]$ using the gaussian quadrature for angular integration.
We calculate the continuum opacity, emissivity and scattering, line opacity and emissivity, mean intensity, collisional rates, and damping coefficient using the RH code.
For the Ca {\sc i} 4227 \AA\, line $\Gamma_R = 2.18\times 10^{8} s^{-1}$. 
\,$\Gamma_E$ is computed using van der Waals broadening (due to elastic collisions with neutral hydrogen) and Stark broadening due to interaction with free electrons \citep[see][]{Uitenbroek_2001}. The micro-turbulent velocity is taken as 3 km/s.

In the second step we use the POLY code \citep[see, e.g.,][]{2005_Holzreuter,Anusha_2011} to solve the two-level atom PRTE by using the quantities obtained from the solution of the RH code kept fixed.
In the second step we only use $80$ wavelength points to cover the Ca {\sc {i}} 4227 \AA\, line, and 6 $\mu$ points in the interval $\mu \in [-1,1]$ using the gaussian quadrature.

\subsection{Computational Considerations}
In this section, we summarize the computational resources used and the time required for the PRTE calculations for a representative column from the 3D rMHD simulation snapshot.
The computations were carried out on the High-Performance Computing (HPC) facility (LEO cluster) at the Indian Institute of Astrophysics, Bengaluru.  
For the present work, we used four nodes, each equipped with 128 cores and 750 GB of RAM. The 1.5D PRTE calculations were carried out by assigning one core to each column, with each column being computed serially.
Both the RH and POLY computations were considered converged when the maximum relative error between successive iterations fell below $10^{-3}$.
For a representative column with 260 depth points, the initial unpolarized multi-level radiative transfer calculation using the RH code requires approximately 25 s.
The subsequent 1.5D polarized radiative transfer calculation with the POLY code requires approximately 17 hours per atmosphere.
The high computational cost arises from the serial nature of the current version of the code, together with the large number of depth points in the 1D columns extracted from 3D rMHD simulations. 
We remark here that the POLY code is currently being parallelized, and we expect a considerable reduction in the computational cost in forthcoming papers. 

\section{Results}
\label{sec:results}
\begin{figure}[htbp]
 \raggedright
\includegraphics[width=1\linewidth]{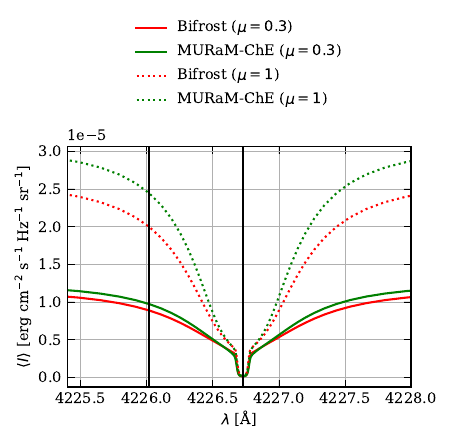}
    \caption{Emergent, spatially averaged intensities for Bifrost (red) and MURaM-ChE (green) models computed at $\mu=1$ (dotted lines) and $\mu=0.3$ (solid lines). Solid black lines show wavelengths at line wing (4226.020 \AA) and at line center (4226.727 \AA) chosen for the discussions (see text in Sect.~\ref{line intensity formation}) 
    }
    \label{fig:intensity}
\end{figure}

%

\subsection{Intensity profiles}
\label{line intensity formation}
To understand the formation of polarization profiles it is important to understand the formation of intensity profiles. 
Fig.~\ref{fig:intensity} shows the emergent, spatially averaged (over $x$ and $y$, denoted by $\langle . \rangle$ ) intensity profiles 
computed along $\mu=1$ (disc center) 
and $\mu=0.3$ (near limb) in Bifrost and MURaM-ChE sampled vertical columns. 
In the line wings, the Bifrost atmospheres produce profiles with lower intensity compared to those arising from MURaM-ChE atmospheres at both $\mu$. 
To understand the emergent intensities we study the formation regions at the line center wavelength $\lambda_\textrm{ref}=4226.727$ \AA\, and a wing wavelength $\lambda_\textrm{ref}=4226.020$ \AA\, marked by vertical lines in Fig.~\ref{fig:intensity}.
\label{line intrnsity formation height}
\begin{figure*}[htbp]
\centering
\includegraphics[width=1\linewidth]{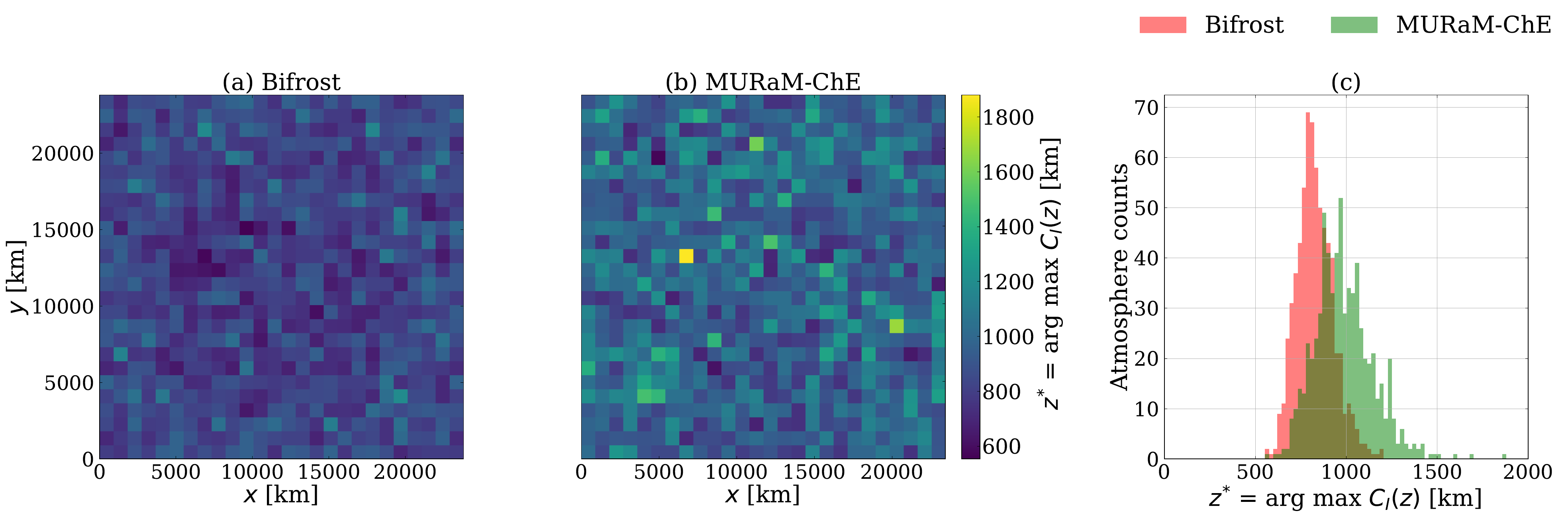}
\caption{Panels (a) and (b) show the heights ($z^{*}$) at which the contribution function $C_{I}$ at the line center wavelength ($\lambda_{\textrm{ref}}=4226.727$ \AA) and at $\mu=1$ is maximum for the sampled slices extracted from Bifrost and MURaM-ChE cubes, respectively. Panel (c) shows the corresponding histograms.} 


\label{fig:4.0.1}
\end{figure*}
\begin{figure*}[htbp]
\centering
\includegraphics[width=0.7\linewidth]{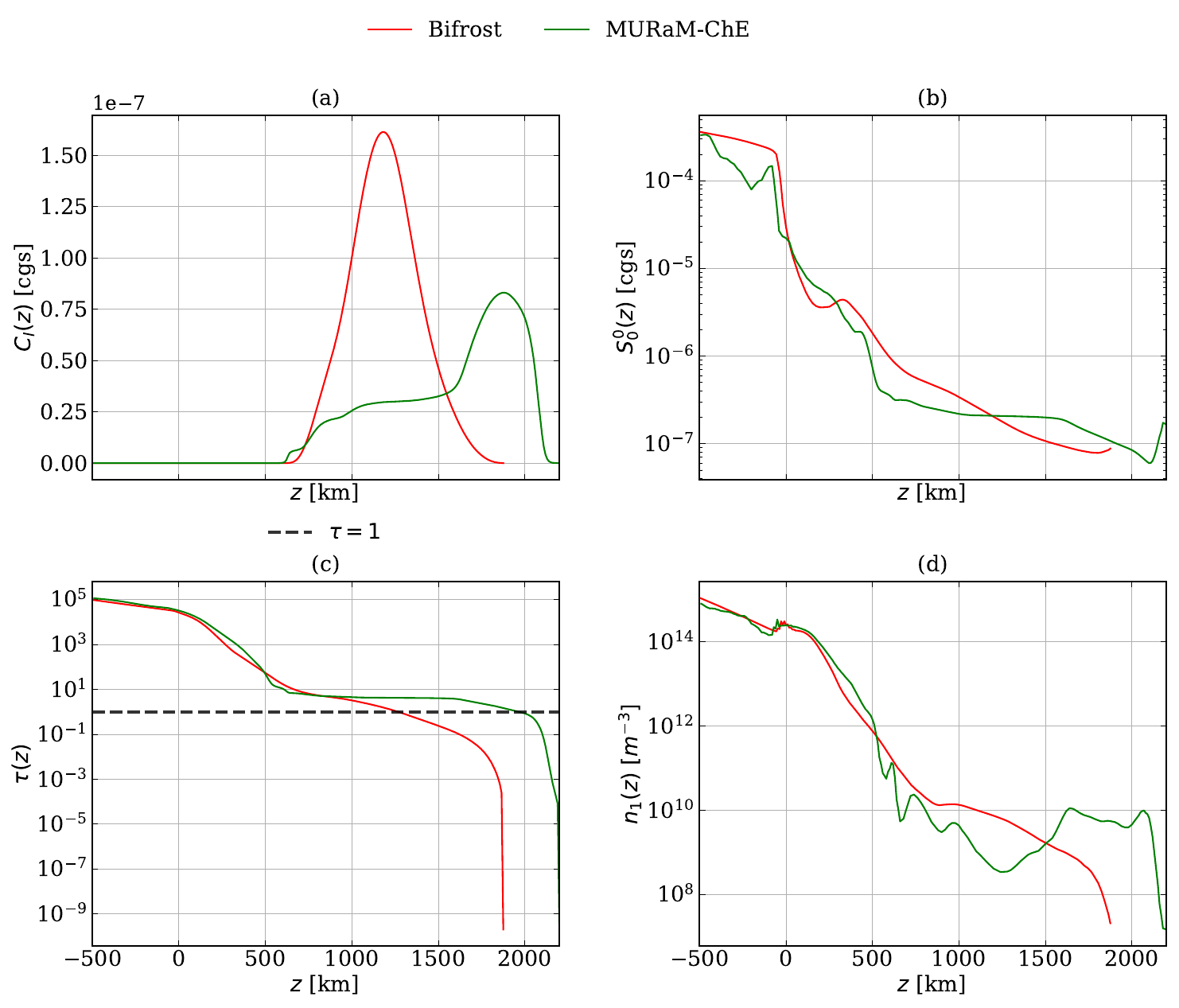}
    \caption{ 
    Panel (a) shows the height variation of the contribution function for Bifrost and MURaM-ChE corresponding to the model atmospheres having maximum height of intensity formation in Fig.~\ref{fig:4.0.1} (a) and (b). Panels (b), (c) and (d) respectively show the height variation of the source function component $S^{0}_{0}$, line-center optical depth, and the ground level population $n_1$ of Ca {\sc{i}} atom for the atmospheres selected to show $C_I$ in panel (a). 
    }
    \label{fig:4.0.1b}
\end{figure*}
In Figs.~\ref{fig:4.0.1}(a) and ~\ref{fig:4.0.1}(b) we compare the heights $z$ where $C_I(\lambda_\textrm{ref}=4226.727$ \AA\,) is maximum along $\mu$ = 1 for the extracted Bifrost and MURaM-ChE vertical columns.
It can be seen that the center of the Ca {\sc i} 4227 \AA\, line is formed over a wider range of heights in MURaM-ChE atmospheres compared to the Bifrost atmospheres. 
This is substantiated by the histograms in Fig.~\ref{fig:4.0.1}(c), which show the distribution of the formation heights displayed in panels (a) and (b). The most probable heights of formation derived from these histograms are, respectively, 969 km and 789 km for MURaM-ChE and Bifrost.  

There are a few outliers in the distribution with large formation heights particularly in MURaM-ChE.
To understand this we now focus on a single column where the height of formation is maximum in Bifrost and MURaM-ChE. 
In Fig.~\ref{fig:4.0.1b}(a) we plot $C_I$ for these two selected atmospheres. 
We observe that the $C_I(\lambda_\textrm{ref}=4226.727$ \AA) is maximum at a greater height for MURaM-ChE ($\approx$ 1880 km) as compared to Bifrost ($\approx$ 1186 km). At these respective heights the amplitude of $C_I(\lambda_\textrm{ref}=4226.727$ \AA) is higher for Bifrost than for MURaM-ChE.
This is because the non-LTE source function component of the intensity denoted by $S^{0}_{0}$ at 1186 km is higher in Bifrost than $S^{0}_{0}$ at 1880 km in MURaM-ChE as seen in Fig.~\ref{fig:4.0.1b}(b). 
The line-center optical depth variation for these two selected atmospheres is plotted in Fig~\ref{fig:4.0.1b}(c), while the height variation of the ground level population density of Ca {\sc i} atom denoted by $n_1$ is shown in Fig.~\ref{fig:4.0.1b}(d).

The dynamics in these two selected atmospheres show that, starting from the top of the atmosphere, the population density in Bifrost gradually increases to the height 1186 km,
while it's rise is steep in MURaM-ChE leading to a greater height (1880 km) at which optical depth reaches unity, which explains the difference in the height of formation of the line center photons in these two atmospheres. 
We also see that there is a dip in $n_1$ variation for MURaM-ChE in the height between 1000 km to 1500 km which results in a decrease in local contribution to the integrated line center opacity. 
Hence the line center optical depth remains nearly constant within the aforementioned height interval for MURaM-ChE.     

\begin{figure}[htbp]
     \centering   \includegraphics[width=1\linewidth]{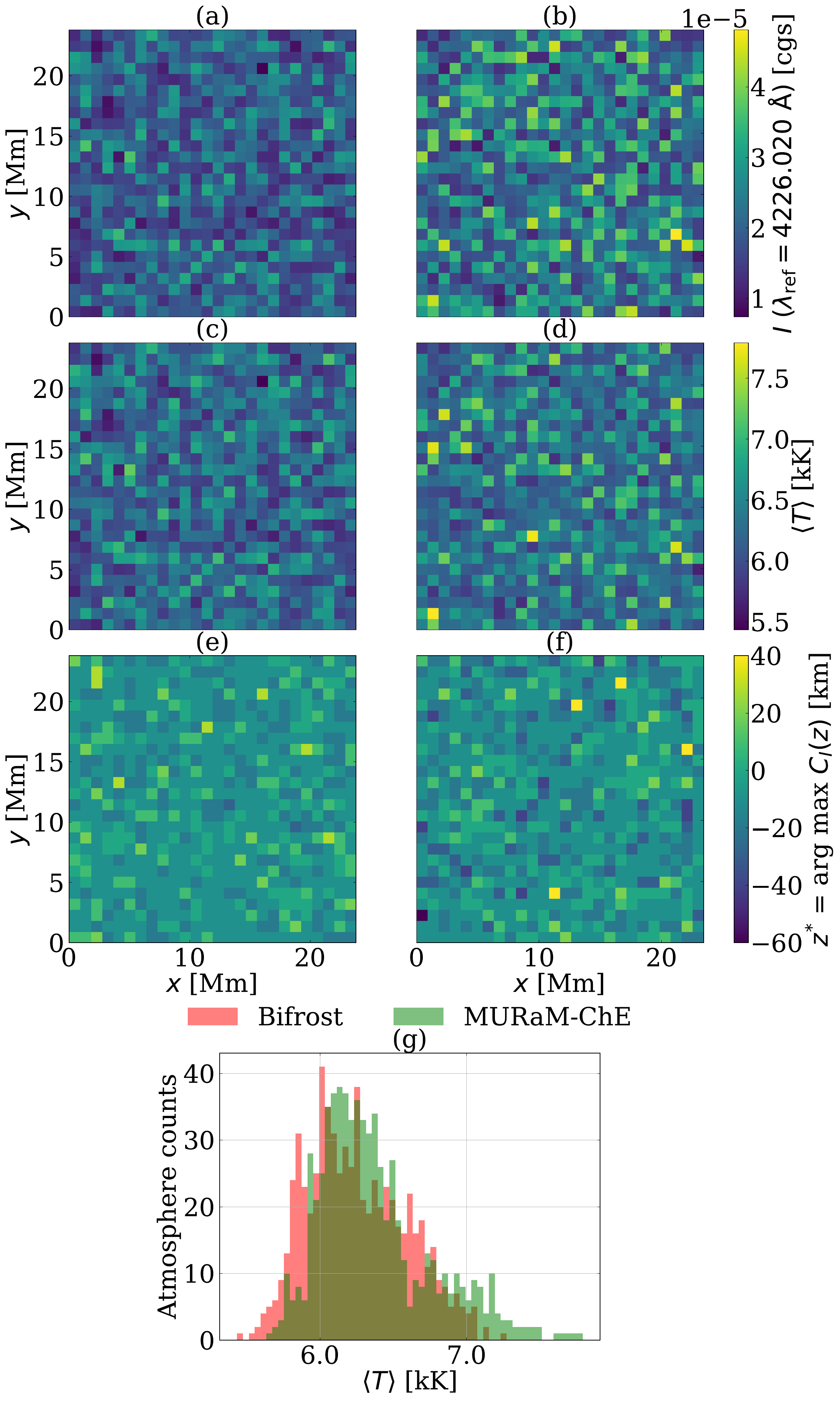}
     \caption{ 
     %
     %
     %
     Panels (a) and (b) show the spatial distribution of the line-wing intensity at $(\lambda_{\mathrm{ref}} = 4226.020\,\AA)$. Panels (c) and (d) show the mean temperature evaluated around $z^{*}=\arg\max C(z)$, while panels (e) and (f) show the corresponding values of $z^{*}$. The first and second columns correspond to the Bifrost and MURaM-ChE models, respectively. Finally, panel (g) shows the histograms of the mean temperatures shown in panels (c) and (d).}
    \label{fig:tempwingmap}
 \end{figure}
%
To understand the formation of the line wings at the wavelength point $\lambda_\textrm{ref} = 4226.020$ \AA\, and along the line of sight $\mu = 1$, we compare the emergent intensity (Fig.~\ref{fig:tempwingmap}(a,b)), the mean temperature around the height $z$ where $C_I$ reaches its maximum (Fig.~\ref{fig:tempwingmap}(c,d)), and the corresponding height $z$ of this maximum contribution (Fig.~\ref{fig:tempwingmap}(e,f)). Panels (a,c,e) correspond to the Bifrost atmosphere, while panels (b,d,f) show the MURaM-ChE atmosphere.
The mean temperature here refers to the temperature averaged over the height range surrounding the peak of $C_I$, corresponding to the region that makes the dominant contribution to the emergent intensity.
We note here that the appearance of noise-like pattern in 2D maps in Fig.~\ref{fig:tempwingmap} is because the structures determining the intensities, temperatures and heights are the granules, which have a size of roughly the horizontal grid resolution of these computations ($\sim$1000 km).
The intensity maps in both simulations correlate with the local temperature maps at the heights of formation in the line wings.
This is because the line wings are formed in the photosphere where local conditions determine the spectral features.
In both simulations the height of wing formation is around $z\sim 0$ km.
In the MURaM-ChE atmospheres, this occurs at slightly lower layers compared to the Bifrost atmospheres.
To substantiate this, in Fig.~\ref{fig:tempwingmap}(g) we show the histograms of the mean temperatures around the heights of formation of the wings in these two simulations.
The distribution of mean temperatures spans a wider range in the MURaM-ChE and is shifted toward higher temperatures, corresponding to lower atmospheric layers, consistent with the photospheric temperature stratification.
The most probable mean temperature in Bifrost is 6013 K which is slightly lower than 6133 K in MURaM-ChE.
Thus the distribution of line-wing formation heights in the MURaM-ChE is shifted toward lower and hotter layers of the photosphere compared to that in the Bifrost, leading to higher wing intensities in the former case.
%
%

However, the emergent, spatially averaged intensity profiles overlap in the line core.
For the Ca {\sc i} 4227 \AA,\ line, the core is formed in the chromosphere, where non-LTE effects become important.
Although the atmospheric structuring differs between the two simulations, with the MURaM-ChE atmospheres exhibiting a higher most probable chromospheric temperature than Bifrost \citep[see Fig. 11 in][]{Ondratschek}, the decoupling of the radiation field from the local temperature structure implies that the emergent intensity is insensitive to these thermal differences.
\subsection{Linear polarization due to scattering in the absence of magnetic fields}
\label{stokes}

%

\begin{figure*}[htbp]
    \centering
    \includegraphics[width=1\linewidth]{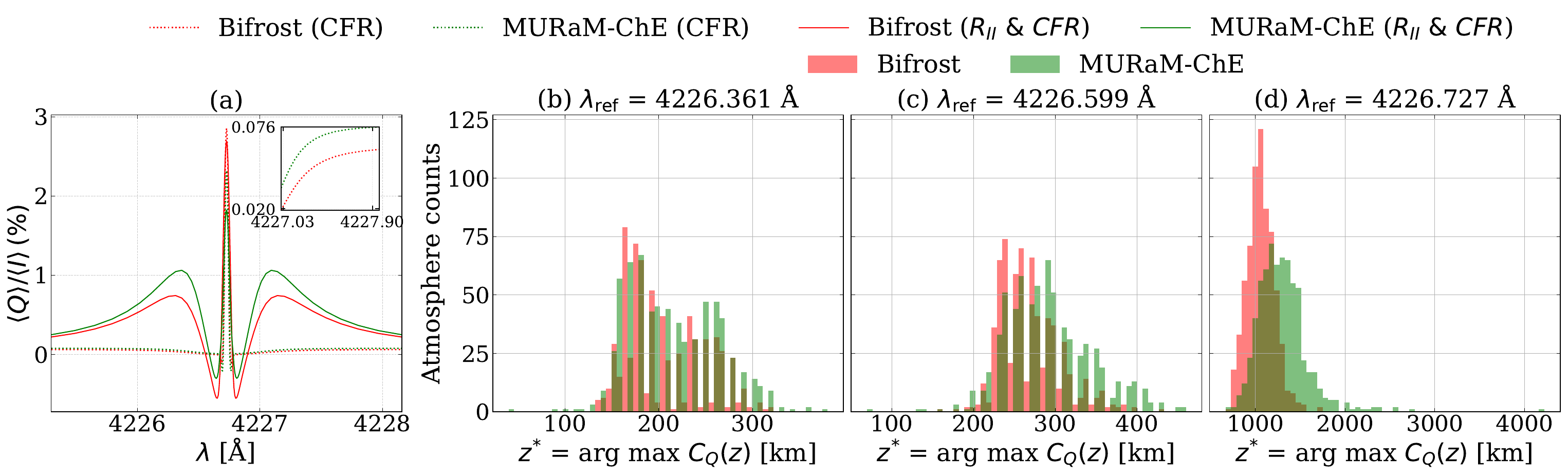}
    \caption{Panel (a) shows the $\langle Q \rangle/\langle I \rangle$ along $\mu=0.3$ in the absence of magnetic fields and for the cases of PFR (solid lines) and CFR (dotted lines). 
    Inset plot highlights weak polarization signature formed in line wing region for the case of CFR. 
    Panels (b), (c) and (d) show histograms of polarization formation height at line wing, line core minima and line center wavelengths, respectively, in the case of PFR. 
    }
    \label{fig:4.1.2}
\end{figure*}
\begin{figure*}[htbp]
    \centering
    \includegraphics[width=1\linewidth]{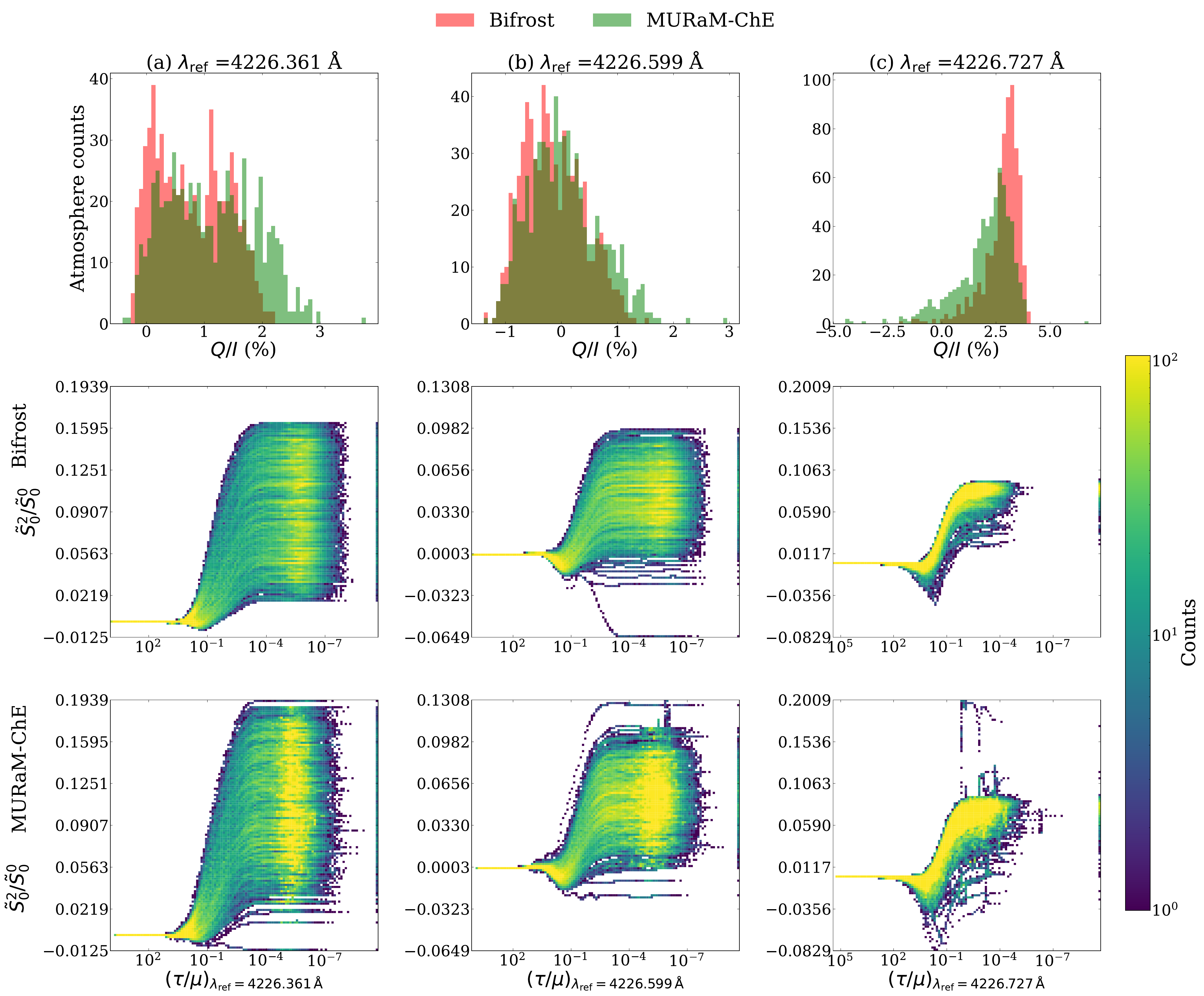}
    \caption{Panels (a), (b) and (c) show the distribution of $Q /I$ (first row) across all the atmospheres, density plot of source function ratio $\Tilde{S}^{2}_{0}/\Tilde{S}^{0}_{0}$ with
    total optical depth $\tau/\mu$ along $\mu=0.3$ for Bifrost (second row) and MURaM-ChE (third row) at line wing (first column), line core minima (second column) and line center (third column) wavelengths in the absence of magnetic fields.}
    

    \label{fig:densityplots}
\end{figure*}

In Fig.~\ref{fig:4.1.2}(a) we compare the emergent, spatially averaged Stokes profiles computed using the PFR and CFR for the nonmagnetic case ($B=0$) along $\mu=0.3$.
Here the $\langle Q\rangle/\langle I \rangle$ is generated due to resonance scattering \citep[see][]{chandrasekhar1950} of the anisotropic radiation by the Ca {\sc i} atoms. 
Azimuthal symmetry of the radiation field in 1D plane parallel geometry leads to $\langle U\rangle /\langle I \rangle = 0$, since Stokes $U$ arises only when this symmetry is broken.
The $\langle Q\rangle/\langle I \rangle$ profiles generated with PFR exhibit the characteristic triple-peak structure \citep[see, e.g.,][]{mf92,2005_Holzreuter}, whereas the CFR profiles show a single peak at the line center with weak signals in the core minima region \citep[see e.g.,][]{2005_Holzreuter} and in the wings (see inset panel of Fig.~\ref{fig:4.1.2}(a)) produced by continuum scattering effects.
The peak value of $\langle Q\rangle/\langle I \rangle$ at the line center in the case of PFR is 2.68\% for Bifrost and 1.82\% for MURaM-ChE.
We can see that at the line center 
the approximation of CFR overestimates the magnitude of $\langle Q \rangle/\langle I \rangle$ in both Bifrost and MURaM-ChE. 
%
%

%
In the core minima region the maximum amplitude of the $\langle Q \rangle/\langle I \rangle$ produced by PFR is $\sim$ 0.30\% in MURaM-ChE which is smaller than $\sim$ 0.55\% produced in Bifrost.
Whereas in the line wings the PFR produces a maximum of $\sim$ 1.06\% $\langle Q \rangle/\langle I \rangle$ in MURaM-ChE which is larger than $\sim$ 0.74\% produced by Bifrost.
%


%
To understand the heights where resonance scattering polarization is generated we compute $C_Q$ as defined by Eq.~(\ref{cq}) in the nonmagnetic case. 
Here we assume for simplicity that Stokes $U$ is generated at the same heights as Stokes $Q$.
We plot histograms of heights where $C_Q$ is maximum, for all the sampled atmospheres in Fig.~\ref{fig:4.1.2}(b), (c) and (d) 
corresponding to three wavelength positions: the line wing at $4226.361$ \AA, the core minima at $4226.599$ \AA\,, and the line center at $4226.727$ \AA.
Since the wavelengths at which the synthetic $Q/I$ profiles attain the wing-peak and core-minimum positions vary among the sampled rMHD model atmospheres, we adopted the reference wavelength positions from the analysis of \citet{2005_Holzreuter}, which was carried out using the FAL-C atmosphere. 
The most probable polarization formation heights obtained from these histograms at line wing, core minima and line center are respectively 164 km (181 km), 240 km (291 km) and 1060 km (1180 km) for Bifrost (MURaM-ChE). 
The differences between heights of polarization formation in the two simulations at line wing, core minimum, and line center positions are respectively 17 km, 51 km, and 120 km.
These differences in heights of polarization formation cause respective differences in $\langle Q \rangle/\langle I \rangle$ amplitudes of 0.32\%, 0.25\%, and 0.86\%.
Thus, scattering polarization is sensitive to even small variations in heights of polarization formation.

To examine the wavelength dependence of the spatially averaged profiles, we plot histograms of $Q / I $ for all the sampled atmospheres in the first row of Fig.~\ref{fig:densityplots} at  the line wing, the core minima and the line center wavelength positions.
%

In the line wing, the range of $Q/I$ values is broader in MURaM-ChE than in Bifrost, while both distributions remain predominantly positive and do not exhibit a single dominant peak.
Consequently, the average $\langle Q \rangle / \langle I \rangle$ is positive in both cases, with a larger amplitude in MURaM-ChE than in Bifrost.

In the core minima region, the distribution of $ Q / I $ is centered around $-0.09\%$ in MURaM-ChE which is closer to zero than -0.32\% around which the Bifrost atmospheres are centered. 
Although the distributions span the range $\sim$ -1.5\% to $+1.5\%$, cancellations during the averaging process significantly reduce the net amplitudes.
As a result of a marginal shift of the distribution toward negative values, Bifrost exhibits a larger absolute amplitude of $\langle Q \rangle / \langle I \rangle$ than MURaM-ChE.

At the line center, the distribution of $ Q  / I $ is largely positive in both simulations. 
In MURaM-ChE the distribution is shifted toward negative values and shows an extended negative tail, and is peaked around 2.68\% whereas in Bifrost it is narrow and sharply peaked around a higher positive value of 3.25\%. 
Consequently, the average $\langle Q \rangle / \langle I \rangle$ is relatively high in the Bifrost.

To understand the distribution of $Q/I$ at a given wavelength,
in the second and third rows of Fig.~\ref{fig:densityplots}, we show, respectively for the Bifrost and MURaM-ChE, the density distributions of the sampled atmospheres with respect to the monochromatic total optical depth (line and  continuum) $\tau_{\lambda_{\mathrm{ref}}}/\mu$ and the source function ratio $\tilde{S}^2_0/\tilde{S}^0_0$ (see Eq.~\eqref{stilde}). 
They exhibit a similar trend at all wavelengths; however, it shifts toward higher optical depths from the line wing to the line center, reflecting the corresponding shift in line formation from lower to higher optical depths. 
%
The source function ratio is controlled by the interplay between angular dependence of the outgoing radiation and that of incoming radiation \citep[see e.g.,][]{2005_Holzreuter}.

In the deeper, optically thick layers of the atmosphere, the radiation field is isotropic, and consequently the anisotropy approaches zero.
At the upper boundary, 
below a critical value of $\tau_{\lambda}/\mu$ the source function ratio as well as the individual components $\tilde{S}^2_0$ and $\tilde{S}^0_0$ (not shown here) become constant as in this region the radiation field decouples from the local conditions. 
This critical optical depth increases from the line wing to the line center, consistent with the increasing opacity toward the line core.
The magnitude and sign of the constant value of the source function ratio and individual components depend on the  atmospheric structure that shapes the radiation field differently across atmospheres and thereby alters the scattering dynamics.
In this part of the atmosphere, due to the constancy of the individual components $\tilde{S}^2_0$ and $\tilde{S}^0_0$, and the fact that the optical depth is small, the resulting irreducible radiation field components\footnote {$\tilde{I}^K_Q $ can be computed using Eqs.~(\ref{ikq-mup}) and (\ref{ikq-mun}) with $\tilde{S}^K_Q$ as the source term. }$\tilde{I}^2_0$ and $\tilde{I}^0_0$ scale with the values of $\tau_{\lambda}/\mu$.
Hence, this regime does not contribute significantly to the formation of the polarization signals (as $Q \propto \tilde{I}^2_0$, $I \propto\tilde{I}^0_0$).
Below this critical optical depth, the source function ratio shows a linear trend, which further takes a nonlinear form with respect to the optical depth scale. 
These regions have intermediate $\tau_{\lambda}/\mu$ values and correspond to the primary formation of polarization signals.

A comparison of the density distribution of atmospheres across wavelengths shows that the  optical depth is smaller in the linear regime than in the nonlinear regime.
In the line wings, the linear regime with positive source function ratio dominates over the nonlinear regime, with the latter concentrated about zero (yellow tint). 
Furthermore, in the linear regime, the distribution spans a broader interval of source function ratio values in MURaM-ChE than in Bifrost, which explains the higher spread of $Q/I$ values obtained in MURaM-ChE (see Fig.~\ref{fig:densityplots} top row, panel a).
In contrast, in the core minima region, the contribution from the nonlinear regime concentrated toward negative values increases, which competes with the contribution from the linear regime spread across the positive values. 
Therefore, the distribution of $Q  / I$ is similar in both Bifrost and MURaM-ChE, centered close to zero (Fig.~\ref{fig:densityplots} top row, panel b). 

At the line center, the density distribution of the atmospheres is concentrated primarily in the positive, linear source function regime.
In the nonlinear regime,  the concentration toward negative values increases in MURaM-ChE, when compared to that in Bifrost.
Therefore, at the line center, the distribution of $ Q  /  I $ is largely concentrated at positive values in Bifrost, whereas it spans over both positive and negative values and exhibits a longer negative tail in MURaM-ChE (Fig.~\ref{fig:densityplots} top row, panel c). We note that the analysis presented here is primarily qualitative. A quantitative investigation would require the use of response functions and will be undertaken in a separate study. \\
Our analysis identifies the physical origin of the differences between the Bifrost and MURaM-ChE results. 
In particular, we find that variations in the height of polarization formation, combined with differences in the  optical structure of the atmosphere at those heights, directly control the amplitude and shape of the emergent scattering polarization profiles. 
Due to the sensitivity of the scattering polarization in the Ca {\sc i} 4227 \AA\ line to the atmospheric parameters such as the formation height,  polarization in this line can serve as a diagnostic for constraining the accuracy of rMHD simulations. 
%
%
\subsection{Impact of Hanle effect}
\label{modelatmos}

\begin{figure}[htbp]
    \centering
    \includegraphics[width=1\linewidth]{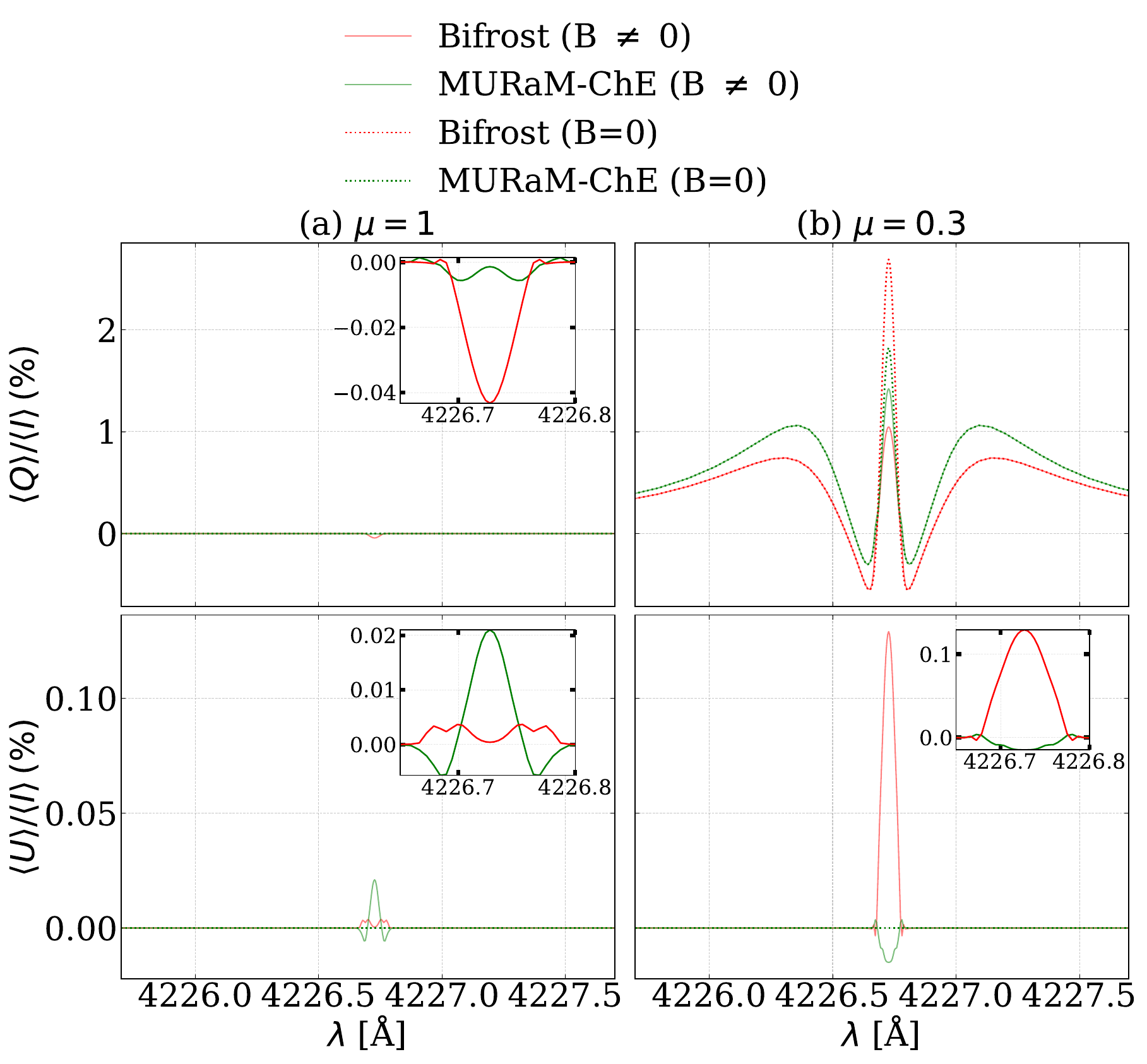}
    \caption{Panels (a) and (b) compare $\langle Q \rangle/\langle I \rangle$ and $\langle U \rangle/\langle I \rangle$ for zero magnetic fields (dotted lines) with those computed using depth dependent magnetic fields (solid lines) along $\mu=1$ and $\mu=0.3$ respectively. Red and green curves represent Bifrost and MURaM-ChE atmospheres respectively.}
    

    \label{fig:4.4.1}
\end{figure}
In this section our aim is to understand how depth dependent magnetic fields in 3D rMHD simulations modify the resonance scattering polarization via the Hanle effect.
For this purpose, in Fig.~\ref{fig:4.4.1} we compare the spatially averaged, emergent Stokes profiles computed using nonmagnetic and magnetic cases for the two 3D rMHD simulation snapshots that we have considered.

In the absence of magnetic and plasma bulk velocity fields, the assumption of a 1D geometry imposes azimuthal symmetry on the radiation field, resulting in $\langle U\rangle/\langle I\rangle = 0$. In the particular case of $\mu=1$, this symmetry also leads to $\langle Q\rangle/\langle I\rangle = 0$.
Introduction of magnetic fields generate a non-zero polarization (see inset panels) reaching a maximum $|\langle Q \rangle/\langle I \rangle|$ ($|\langle U \rangle/\langle I \rangle|$ ) of 0.0432\% (0.0036\%) and 0.0056\% (0.0209\%) for Bifrost and MURaM-ChE respectively.

Along $\mu=0.3$ we clearly observe Hanle depolarization in $\langle Q \rangle/\langle I \rangle$ for both Bifrost and MURaM-ChE with the magnitude of depolarization being higher in Bifrost (1.643\%) when compared to MURaM-ChE (0.401\%). 
Further a maximum $\langle U \rangle/\langle I \rangle$ amplitude of 0.129\% (0.015\%) is generated in Bifrost (MURaM-ChE).


\begin{figure*}[htbp]
     \centering   \includegraphics[width=1\linewidth]{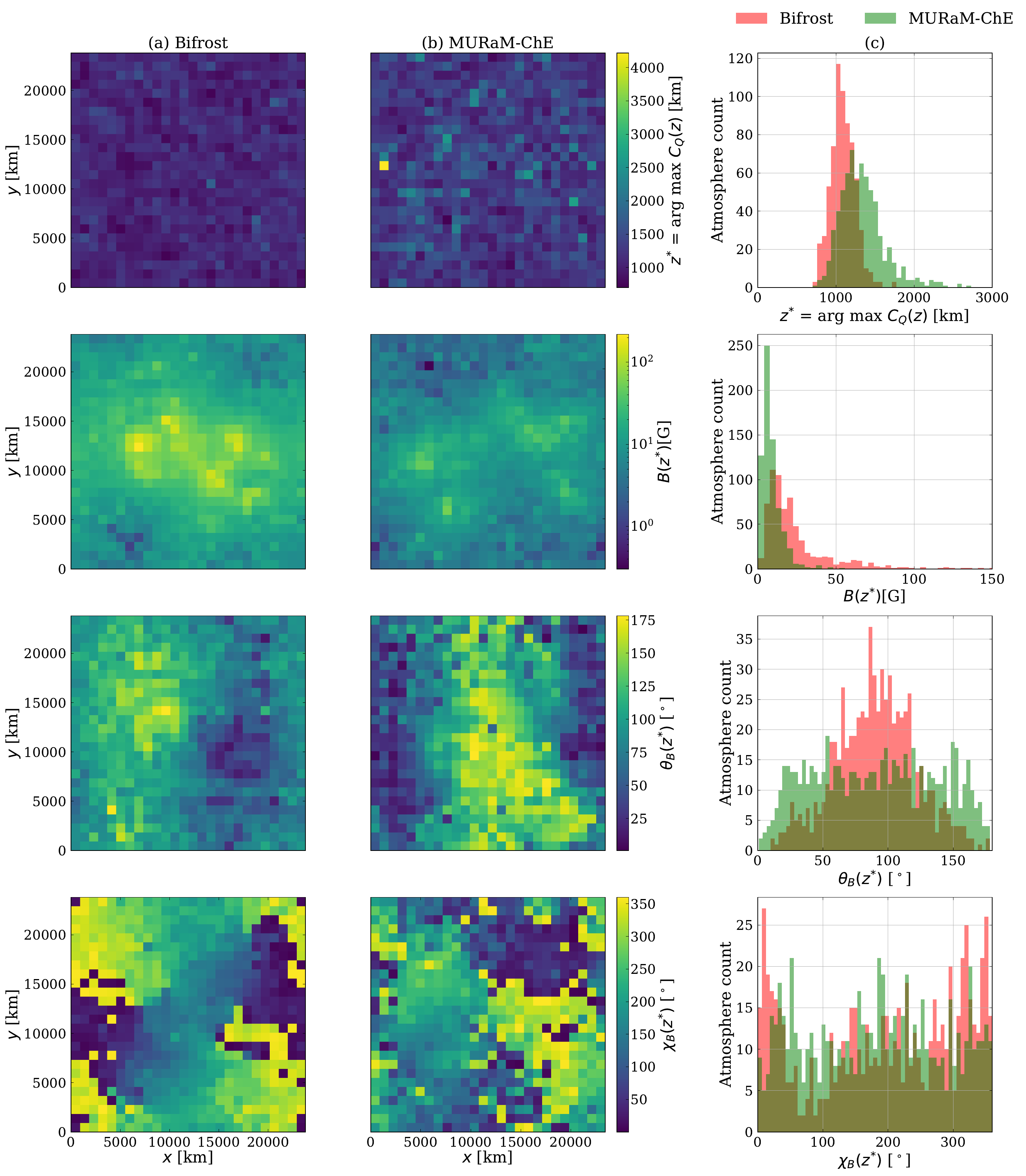}
\caption{Columns (a) and (b) show the spatial distributions of the vertical height (first row), strength B (second row),
polar angle (third row), and azimuthal angle (fourth row) of the magnetic field vector, evaluated at the height where the contribution function $C_Q$ reaches its maximum (at line center wavelength, $\lambda_{\textrm{ref}} = 4226.727\,\AA$), for Bifrost and MURaM-ChE, respectively. Column (c) shows the corresponding histograms.}


     \label{fig:bmap}
 \end{figure*}

To identify the magnetic field distribution at the heights where polarization is generated, the contribution function $C_Q$ in the magnetic case is computed.
The 2D maps of heights $z$ at which $C_Q$ attains its maximum are presented in the first row of Fig.~\ref{fig:bmap}(a) and (b), while the corresponding histograms of these heights are displayed in Fig.~\ref{fig:bmap}(c).
%
The most probable heights of polarization for Bifrost and MURaM-ChE are 1028 km and 1207 km respectively. 
Clearly, line core polarization is formed higher in MURaM-ChE than compared with Bifrost consistent with formation heights of line core intensity.
The maxima of $C_Q$ occur at greater heights than those of $C_I$, since scattering polarization forms higher in the atmosphere than intensity \citep[see, e.g.,][]{2005_Holzreuter, Mathur_2025}.

The 2D maps of the vector magnetic field at the heights where the line core polarization is formed are shown in the second, third, and fourth rows of Figure~\ref{fig:bmap}(a) and (b). 
The magnetic field strength distributions at the polarization formation height exhibit maximum values of 54.9 G and 218.6 G for the selected MURaM-ChE and Bifrost model atmospheres, respectively, because the Ca {\sc i} 4227 \AA\, line-core polarization forms at greater depths in Bifrost than in MURaM-ChE.
The histogram in the second row of panel (c) further shows that the most probable magnetic field strengths are 9.5 G in Bifrost and 5.8 G in MURaM-ChE, consistent with the higher formation height of the line-core polarization in MURaM-ChE, where the magnetic field is weaker.
%
%
We see that the spread of the weak field distribution is higher in Bifrost leading to a higher depolarization of $\langle Q \rangle/\langle I \rangle$ than that in MURaM-ChE.

The polar and azimuth angle maps are shown in third and fourth rows in panels (a) and (b) of Fig.~\ref{fig:bmap}. 
It is clear that in MURaM-ChE the magnetic field inclinations are nearly uniformly distributed  while in Bifrost they are more clustered around $90 \deg$.

This is also substantiated by the histograms in Fig.~\ref{fig:bmap}(c) third row. 
The field azimuths are almost uniformly distributed in both Bifrost and MURaM-ChE atmospheres.
Because of this, the spatially  averaged $\langle U\rangle/\langle I \rangle$ has a relatively smaller amplitude compared to that of $\langle Q\rangle/\langle I \rangle$. 
%
%
We note that, the differences in magnetic field distribution in the two simulations at the height of polarization formation is largely a consequence of the differences in the formation height of line-core polarization, rather than a fundamental difference in the underlying magnetic structures.
\section{Conclusions}
\label{sec:conclusions}
In this paper we study the emergent, spatially averaged polarization spectra of Ca {\sc{i}} 4227 \AA\, line obtained by solving 1.5D PRTE with angle-averaged PFR and magnetic fields using 1D vertical columns extracted from snapshots of 3D rMHD Bifrost and MURaM-ChE simulations.
Plasma bulk velocity fields are ignored in this work. 
%

A comparison of the spatially averaged emergent intensity profiles from the two simulations shows similar line cores, while MURaM-ChE exhibits a higher line wing and continuum intensity. 
The heights at which the contribution function peaks indicate that the line core forms at greater heights in MURaM-ChE atmospheres than in Bifrost ones. 
However, the similarity of the core intensities suggests that they are largely insensitive to the differences in formation height between the two simulations.
In contrast, the line wings form slightly higher in Bifrost, accounting for its lower wing intensity. 
This reflects differences in the thermal structure of the line-wing formation regions. In particular, the most probable temperature is higher in MURaM-ChE than in Bifrost.

%
%
Comparison of nonmagnetic, emergent, spatially averaged Stokes spectra shows that Bifrost atmospheric models produce higher magnitude of resonance scattering $\langle Q \rangle/\langle I \rangle$ at line center and line core minima but lower value in line wing as compared with those from MURaM-ChE. 
The distributions of $Q/I$ in both Bifrost and MURaM-ChE atmospheric models do not exhibit a single dominant peak at line wings, while MURaM-ChE spans over a wider positive range of $Q/I$, resulting in a higher spatially averaged value than in Bifrost at this wavelength.
At the core minima wavelength, both simulations show that the distributions of $Q/I$ is centered at around zero, with MURaM-ChE being closer to zero than Bifrost. 
This results in a lower magnitude of emergent, spatially averaged polarization, $\langle Q \rangle / \langle I \rangle$, for MURaM-ChE.
At line center, Bifrost shows a higher average $\langle Q \rangle / \langle I \rangle$ than MURaM-ChE due to its narrow distribution peaked at a slightly higher value than MURaM-ChE, with MURaM-ChE shifted toward lower values with a negative tail.
To further investigate the behavior, we analyzed the depth dependence of the source function ratio  $\tilde{S}^2_0/\tilde{S}^0_0$.
This shows that $\langle Q \rangle/\langle I \rangle$ is mainly generated in the intermediate optical depth regime, where $\tilde{S}^2_0/\tilde{S}^0_0$ is initially linear and then becomes non-linear with increasing optical depth.
A more quantitative analysis of the sensitivity of scattering polarization to various parameters require response functions which we will address in a future study.

In the presence of depth dependent magnetic fields, 
we observe that Bifrost atmospheres undergo higher Hanle depolarization in  $\langle Q \rangle / \langle I \rangle$ than MURaM-ChE. 
%
%
This difference arises because the distribution of magnetic field strengths in Bifrost is skewed, spanning a broad range of 0–100 G and peaking near 10 G in the line-forming region, whereas MURaM-ChE exhibits a much narrower distribution, with a peak around 6 G. 
%
The line core polarization in MURaM-ChE is formed at higher region with weaker magnetic fields relative to that in Bifrost. 
This leads to stronger line-core depolarization in Bifrost than in MURaM-ChE. 
A meaningful comparison with observations, however, requires the inclusion of plasma bulk velocity fields, and a more realistic treatment of scattering using angle-dependent PFR and 3D PRTE.
Modeling the observed profiles with one or more of these additional physical effects will be addressed in forthcoming papers.
\begin{acknowledgements}
DA and LSA acknowledge financial support from the Science and Engineering Research
Board (SERB)/Anusandhan National Research Foundation
(ANRF), India, under the Core Research Grant (CRG), Grant
No. CRG/2022/001522. This research has made use of the
High-Performance Computing (HPC) resources (LEO cluster) made available by the Computer Center of the Indian Institute of Astrophysics, Bengaluru.\\
Software: NumPy \citep{2020_Harris}, matplotlib
\citep{2007_Hunter}, RH \citep{Uitenbroek_2001}, POLY
\citep{2005_Holzreuter,Anusha_2011}.
    
\end{acknowledgements}

\bibliographystyle{aa} 
\bibliography{refs}

\end{document}